\documentclass{JHEP3}
\usepackage[dvips]{graphicx}
\usepackage[applemac]{inputenc}

\title{On the gluon content of the {\boldmath $\eta$} and {\boldmath $\eta^\prime$} mesons}
\author{
Rafel Escribano\\
Grup de Física Teòrica and IFAE,
Universitat Autònoma de Barcelona,
E-08193 Bellaterra (Barcelona), Spain\\
E-mail: \email{Rafel.Escribano@ifae.es}
}
\author{
Jordi Nadal\\
Institut de Física d'Altes Energies,
Universitat Autònoma de Barcelona,
E-08193 Bellaterra (Barcelona), Spain\\
E-mail: \email{jnadal@ifae.es}
}
\abstract{
A phenomenological analysis of radiative $V\to P\gamma$ and $P\to V\gamma$ decays is performed
with the purpose of determining the gluonic content of the $\eta$ and $\eta^\prime$ wave functions.
Our results show that within our model there is no evidence for a gluonium contribution in the $\eta$,
$Z_\eta^2=0.00\pm 0.12$, or the $\eta^\prime$, $Z_{\eta^\prime}^2=0.04\pm 0.09$.
In terms of a mixing angle description this corresponds to $\phi_P=(41.4\pm 1.3)^\circ$ and
$|\phi_{\eta^\prime G}|=(12\pm 13)^\circ$.
In addition, the $\eta$-$\eta^\prime$ mixing angle is found to be
$\phi_P=(41.5\pm 1.2)^\circ$ if we don't allow for a gluonium component.}
\keywords{pmo, emp}
\preprint{UAB--FT--628}

\begin{document}
\section{Introduction}
\label{intro}
Very recently, the KLOE Collaboration has reported a new measurement of the ratio
$R_\phi\equiv B(\phi\to\eta^\prime\gamma)/B(\phi\to\eta\gamma)$ \cite{Ambrosino:2006gk}.
Combining the value of $R_\phi$ with other constraints,
they have estimated the gluonium content of the $\eta^\prime$ meson as
$Z_{\eta^\prime}^2=0.14\pm 0.04$,
which points to a significant gluonium fraction in the $\eta^\prime$ wave function
incompatible with zero by more than $3\sigma$.
This new result contrasts with the former value $Z_{\eta^\prime}^2=0.06^{+0.09}_{-0.06}$,
which was compatible with zero within $1\sigma$ and consistent with a gluonium fraction below 15\%
\cite{Aloisio:2002vm}.
Both experimental analysis obtain their results using the same set of constraints derived from other measured ratios, namely\footnote{
The analysis in Ref.~\cite{Aloisio:2002vm} did not contain the constraint imposed by
$\Gamma(\eta^\prime\to\omega\gamma)/\Gamma(\omega\to\pi^0\gamma)$.
Nevertheless, the inclusion of this constraint does not change the main results of the analysis,
as seen in Ref.~\cite{Ambrosino:2006gk}.}
$\Gamma(\eta^\prime\to\gamma\gamma)/\Gamma(\pi^0\to\gamma\gamma)$,
$\Gamma(\eta^\prime\to\rho\gamma)/\Gamma(\omega\to\pi^0\gamma)$ and
$\Gamma(\eta^\prime\to\omega\gamma)/\Gamma(\omega\to\pi^0\gamma)$, as well as $R_\phi$.
The sole difference between the two analyses, apart from the obvious improvement in the precision of the new measurements, is the inclusion in the amplitudes of
Ref.~\cite{Ambrosino:2006gk}
of two extra parameters to deal with the overlap of the vector and pseudoscalar meson wave functions produced in the transitions  $V\to P\gamma$ or $P\to V\gamma$, a feature first introduced in
Ref.~\cite{Bramon:2000fr}.
However, the new analysis of Ref.~\cite{Ambrosino:2006gk} uses the most recent experimental data taken from Ref.~\cite{Yao:2006px} in association with the values for the parameters related to the overlap which were obtained in Ref.~\cite{Bramon:2000fr} from a fit to available experimental data at that time.
Therefore, a reanalysis of this uncomfortable situation is needed before drawing definite conclusions on the gluon content of the $\eta$ and $\eta^\prime$ mesons.
The purpose of this work is to perform a phenomenological analysis of radiative
$V\to P\gamma$ and $P\to V\gamma$ decays,
with $V=\rho, K^\ast, \omega, \phi$ and $P=\pi, K, \eta, \eta^\prime$,
aimed at determining the gluonic content of the $\eta$ and $\eta^\prime$ wave functions.
Similar analysis were driven in the seminal work by Rosner \cite{Rosner:1982ey},
where the allowed gluonic admixture in the $\eta$ was found to be small, $|Z_\eta|\lesssim 0.4$,
but not necessarily zero.
The gluonic content of the $\eta^\prime$ could not be established due to the lack of data on
$\phi\to\eta^\prime\gamma$.
Later on, Kou pointed out that the $\eta^\prime$ gluonic component might be as large as
26\% \cite{Kou:1999tt}.
These two works also included in their analysis the constraints provided by
$P\to\gamma\gamma$ with $P=\eta, \eta^\prime$.
Here, we will not take into account these decays since they are nicely described in a
two mixing angle scheme, as shown in Refs.~\cite{Escribano:2005qq,Feldmann:1999uf}.
A more recent analysis studying the pseudoscalar glueball-$q\bar q$ mixing from
$J/\psi(\psi^\prime)\to VP$ decays gets
$|Z_\eta|=0.042$ and $|Z_{\eta^\prime}|=0.161$ \cite{Li:2007ky}.

In Sec.~\ref{notation}, we briefly introduce the notation.
The model for $VP\gamma$ $M1$ transitions to be used in our analysis
is presented in Sec.~\ref{model}.
This model includes the overlap of vector and pseudoscalar wave functions as a main
ingredient \cite{Bramon:2000fr}.
As stated above, the use of old values for the overlap parameters in addition to
the latest data may be at the origin of the discrepancy concerning the gluon content of the $\eta^\prime$ between the two KLOE analyses \cite{Ambrosino:2006gk,Aloisio:2002vm}.
Sec.~\ref{datafit} is devoted to the data fitting of the most recent $V\to P\gamma$ and $P\to V\gamma$ experimental data with the aim of finding the gluonic admixture in the $\eta$ and $\eta^\prime$
wave functions.
As a matter of comparison with other approaches, a graphical representation of our main results is shown in Sec.~\ref{approaches}.
Finally, in Sec.~\ref{conclusions} we summarize our work and present the conclusions.
A decomposition of the mixing parameters in terms of Euler angles is left for App.~\ref{decomposition}.

\section{Notation}
\label{notation}
We will work in a basis consisting of the states
 $|\eta_q\rangle\equiv\frac{1}{\sqrt{2}}|u\bar u+d\bar d\rangle$, $|\eta_s\rangle=|s\bar s\rangle$
and $|G\rangle\equiv|\mbox{gluonium}\rangle$.
The physical states $\eta$ and $\eta^\prime$ are assumed to be linear combinations of these:
\begin{equation}
\label{mixingP}
\begin{array}{rcl}
|\eta\rangle &=& X_\eta|\eta_q\rangle+Y_\eta|\eta_s\rangle+Z_\eta|G\rangle\ ,\\[1ex]
|\eta^\prime\rangle &=&
X_{\eta^\prime}|\eta_q\rangle+Y_{\eta^\prime}|\eta_s\rangle+Z_{\eta^\prime}|G\rangle\ ,
\end{array}
\end{equation}
with $X_{\eta(\eta^\prime)}^2+Y_{\eta(\eta^\prime)}^2+Z_{\eta(\eta^\prime)}^2=1$
and thus $X_{\eta(\eta^\prime)}^2+Y_{\eta(\eta^\prime)}^2\leq 1$.
A significant gluonic admixture in a state is possible only if
$Z_{\eta(\eta^\prime)}^2=1-X_{\eta(\eta^\prime)}^2-Y_{\eta(\eta^\prime)}^2>0$ \cite{Rosner:1982ey}.
This mixing scheme assumes isospin symmetry, \emph{i.e.~}no mixing with $\pi^0$,
and neglects other possible admixtures from $c\bar c$ states and/or radial excitations.
In absence of gluonium, $Z_{\eta(\eta^\prime)}\equiv 0$, the mixing parametrization (\ref{mixingP})
is reduced to the standard pattern in the quark-flavour basis
\begin{equation}
\label{mixingPstandard}
\begin{array}{rcl}
|\eta\rangle &=& \cos\phi_P|\eta_q\rangle-\sin\phi_P|\eta_s\rangle\ ,\\[1ex]
|\eta^\prime\rangle &=&
\sin\phi_P|\eta_q\rangle+\cos\phi_P|\eta_s\rangle\ ,
\end{array}
\end{equation}
with $X_\eta=Y_{\eta^\prime}\equiv\cos\phi_P$, $X_{\eta^\prime}=-Y_\eta\equiv\sin\phi_P$,
and $X_{\eta(\eta^\prime)}^2+Y_{\eta(\eta^\prime)}^2=1$.
Similarly, for the vector states $\omega$ and $\phi$ the mixing is given by
\begin{equation}
\label{mixingV}
\begin{array}{rcl}
|\omega\rangle &=& \cos\phi_V|\omega_q\rangle-\sin\phi_V|\phi_s\rangle\ ,\\[1ex]
|\phi\rangle &=& \sin\phi_V|\omega_q\rangle+\cos\phi_V|\phi_s\rangle\ ,
\end{array}
\end{equation}
where $|\omega_q\rangle$ and $|\phi_s\rangle$ are the analog non-strange and strange states of
$|\eta_q\rangle$ and $|\eta_s\rangle$, respectively.

\section{A model for {\boldmath $VP\gamma$ $M1$} transitions}
\label{model}
We will work in a conventional quark model context and assume that pseudoscalar and vector mesons are simple quark-antiquark $S$-wave bound states.  
All these hadrons are thus extended objects with characteristic spatial extensions fixed by their respective quark-antiquark $P$ or $V$ wave functions.  
In the pseudoscalar nonet, $P=\pi, K, \eta, \eta^\prime$, 
the quark spins are antiparallel and the mixing pattern is given by Eq.~(\ref{mixingP}). 
In the vector case, $V=\rho, K^\ast, \omega, \phi$, the spins are parallel and mixing is 
similarly given by Eq.~(\ref{mixingV}). 
We will work in the good $SU(2)$ limit with $m_u = m_d \equiv\bar{m}$
and with identical spatial extension of wave functions within each $P$ and each $V$ isomultiplet.
$SU(3)$ will be broken in the usual manner taking constituent quark masses with
$m_s > \bar{m}$ but also, and this is a specific feature of our approach, 
allowing for different spatial extensions for each $P$ and $V$ isomultiplet.
Finally, we will consider that even if gluon annihilation channels may induce 
$\eta$-$\eta^\prime$ mixing, they play a negligible r\^ole in $VP\gamma$ transitions and
thus fully respect the usual OZI-rule. 

In our specific case of $VP\gamma$ $M1$ transitions, these generic statements translate 
into three characteristic ingredients of the model: 
\begin{itemize}
\item[{\it i)}] 
A $VP\gamma$ magnetic dipole transition proceeds via quark or antiquark spin-flip 
amplitudes proportional to $\mu_q=e_q/2m_q$. 
Apart from the obvious quark charge values, this effective magnetic moment breaks
$SU(3)$ in a well defined way and distinguishes photon emission from strange or non-strange quarks via $m_s > \bar{m}$.
\item[{\it ii)}] 
The spin-flip $V\leftrightarrow P$ conversion amplitude has then to be corrected by the 
relative overlap between the $P$ and $V$ wave functions. 
In older papers \cite{Isgur:1976hg,O'Donnell:1981sj}
a common, flavour-independent overlap was introduced.
Today, with a wider set of improved data, this new symmetry breaking mechanism can be introduced 
without enlarging excessively the number of free parameters. 
\item[{\it iii)}] 
Indeed, the OZI-rule reduces considerably the possible transitions and their respective 
$VP$ wave-function overlaps: $C_s$, $C_q$ and $C_\pi$ characterize the 
$\langle\eta_s|\phi_s\rangle$, 
$\langle\eta_q|\omega_q\rangle =\langle\eta_q|\rho\rangle$ and 
$\langle\pi|\omega_q\rangle =\langle\pi|\rho\rangle$ spatial overlaps, respectively.
Notice that distinction is made between the $|\pi\rangle$ and $|\eta_q\rangle$ 
spatial extension due to the gluon or $U(1)_A$ anomaly affecting the second state.
Independently, we will also need $C_K$ for the $\langle K|K^\ast\rangle$ overlap
between strange isodoublets. 
\end{itemize}

It is then a trivial task to write all the $VP\gamma$ couplings in terms of an
effective $g\equiv g_{\omega_q\pi\gamma}$:
\begin{equation}
\label{piKcouplings}
\begin{array}{c}
g_{\rho^0\pi^0\gamma}=g_{\rho^+\pi^+\gamma}=\frac{1}{3}g\ ,\quad
g_{\omega\pi\gamma}=g\cos\phi_V\ ,\quad
g_{\phi\pi\gamma}=g\sin\phi_V\ ,\\[2ex]
g_{K^{*0}K^0\gamma}=-\frac{1}{3}g\,z_K\left(1+\frac{\bar{m}}{m_s}\right)\ ,\quad
g_{K^{*+}K^+\gamma}=\frac{1}{3}g\,z_K\left(2-\frac{\bar{m}}{m_s}\right)\ ,\\[2ex]
g_{\rho\eta\gamma}=g\,z_{q}\,X_\eta\ ,\quad
g_{\rho\eta^\prime\gamma}=g\,z_{q}\,X_{\eta^\prime}\ ,\\[2ex]
\end{array}
\end{equation}
\begin{equation}
\label{etacouplings}
\begin{array}{c}
g_{\omega\eta\gamma}=
\frac{1}{3}g\left(z_q\,X_\eta\cos\phi_V +2\frac{\bar{m}}{m_s}z_s\,Y_\eta\sin\phi_V\right)\ ,\\[2ex]
g_{\omega\eta^\prime\gamma}=
\frac{1}{3}g\left(z_q\,X_{\eta^\prime}\cos\phi_V
                           +2\frac{\bar{m}}{m_s}z_s\,Y_{\eta^\prime}\sin\phi_V\right)\ ,\\[2ex]
g_{\phi\eta\gamma}=
\frac{1}{3}g\left(z_q\,X_\eta\sin\phi_V -2\frac{\bar{m}}{m_s}z_s\,Y_\eta\cos\phi_V\right)\ ,\\[2ex]
g_{\phi\eta^\prime\gamma}=
\frac{1}{3}g\left(z_q\,X_{\eta^\prime}\sin\phi_V
                           -2\frac{\bar{m}}{m_s}z_s\,Y_{\eta^\prime}\cos\phi_V\right)\ ,\\[2ex]
\end{array}
\end{equation}
where we have redefined $z_q\equiv C_q/C_\pi$, $z_s\equiv C_s/C_\pi$ and $z_K\equiv C_K/C_\pi$.
The normalization of the couplings is such that 
$g_{\omega\pi\gamma}=g\,\cos\phi_V =
2\,(\mu_u + \mu_{\bar{d}})\,C_\pi\cos \phi_V =e\,C_\pi\cos\phi_V/\bar{m}$ 
and the decay widths are given by 
\begin{equation}
\label{width}
\Gamma (V\rightarrow P\gamma)=
\frac{1}{3}\frac{g^2_{VP\gamma}}{4\pi}|{\bf p}_\gamma|^3\ ,\qquad
\Gamma (P\rightarrow V\gamma)=\frac{g^2_{VP\gamma}}{4\pi}|{\bf p}_\gamma|^3\ , 
\end{equation}
where ${\bf p}_\gamma$ is the final photon momentum.

\section{Data fitting}
\label{datafit}
We proceed to fit our theoretical expressions for the amplitudes 
in Eqs.~(\ref{piKcouplings},\ref{etacouplings}) comparing the available experimental information on 
$\Gamma (V\rightarrow P\gamma)$ and $\Gamma (P\rightarrow V\gamma)$
taken exclusively from Ref.~\cite{Yao:2006px} with the corresponding decay widths in Eq.~(\ref{width}).
Looking at these amplitudes, one immediately realizes that the overlapping parameters $z_{q,s}$ and the mixing parameters $X_{\eta,(\eta^\prime)}$ and $Y_{\eta,(\eta^\prime)}$ always appear in pairs,
namely, $z_q X_{\eta,(\eta^\prime)}$ and $z_s Y_{\eta,(\eta^\prime)}$.
Constraining these four different combinations will fix four parameters at most.
However, there are five independent parameters to be fixed from them,
viz.~$z_{q,s}$ and the three mixing parameters related to the most general case of accepting a gluonic admixture in both the $\eta$ and $\eta^\prime$ mesons (see below).
So, either we fix the $z$'s to unity and then constrain the three mixing parameters of the general case,
or we leave the $z$'s free and then we are restricted to allow for gluonium either in the
$\eta$ \emph{or} $\eta^\prime$ wave function only.
In the following, we consider these possibilities.
Both are interesting since a comparison of their results will allow us to check the relevance of taking into consideration the overlapping parameters, which are specific of our approach.
Furthermore, leaving the $z$'s free will permit us to fix the gluonic content of the $\eta^\prime$ in a way identical to the experimental measurement by KLOE, that is, under the hypothesis of no gluonium in the $\eta$ wave function.
Unfortunately, due to the pairing of parameters mentioned above, a simultaneous fit of the $z$'s and the gluonic admixture in the $\eta$ \emph{and} $\eta^\prime$ is not possible.

As shown in detail in App.~\ref{decomposition}, due to the orthonormality conditions in
Eqs.~(\ref{normalization},\ref{orthogonality}) the mixing pattern of $\eta$ and $\eta^\prime$ is described by means of three independent parameters, which we choose to be
$X_\eta$, $X_{\eta^\prime}$ and $Y_{\eta^\prime}$.
Accordingly, the parameters related to the gluonic content, $Z_\eta$ and $Z_{\eta^\prime}$,
and $Y_\eta$ are written in terms of the former ones as
\begin{equation}
\label{Zs}
\begin{array}{c}
|Z_{\eta,(\eta^\prime)}|=\sqrt{1-X_{\eta,(\eta^\prime)}^2-Y_{\eta,(\eta^\prime)}^2}\ ,\\[2ex]
Y_\eta=-\frac{\displaystyle X_\eta X_{\eta^\prime}Y_{\eta^\prime}+
                        \sqrt{(1-X_{\eta^\prime}^2-Y_{\eta^\prime}^2)(1-X_\eta^2-X_{\eta^\prime}^2)}}
                      {\displaystyle 1-X_{\eta^\prime}^2}\ .
\end{array}
\end{equation}
The mixing parameters can also be expressed in terms of three angles,
$\phi_P$, $\phi_{\eta G}$ and $\phi_{\eta^\prime G}$, the two latter weighting the gluonic admixture in the $\eta$ and $\eta^\prime$, respectively.
Using this angular parametrization one gets
\begin{equation}
\label{angular}
\begin{array}{c}
X_\eta=\cos\phi_P\cos\phi_{\eta G}\ ,\quad
X_{\eta^\prime}=\sin\phi_P\cos\phi_{\eta^\prime G}-
                              \cos\phi_P\sin\phi_{\eta G}\sin\phi_{\eta^\prime G}\ ,\\[2ex]
Y_\eta=-\sin\phi_P\cos\phi_{\eta G}\ ,\quad
Y_{\eta^\prime}=\cos\phi_P\cos\phi_{\eta^\prime G}+
                              \sin\phi_P\sin\phi_{\eta G}\sin\phi_{\eta^\prime G}\ ,\\[2ex]
Z_\eta=-\sin\phi_{\eta G}\ ,\quad
Z_{\eta^\prime}=-\sin\phi_{\eta^\prime G}\cos\phi_{\eta G}\ .
\end{array}
\end{equation}
Fits to experimental data are performed imposing the constraints in Eq.~(\ref{Zs}) or, equivalently,
using the decomposition in Eq.~(\ref{angular}).

We start considering the first of the possibilities noted before.
Thus, we assume that the overlap of the $P$ and $V$ wave functions is flavour-independent,
\textit{i.e.}~$C_q=C_s=C_K=C_\pi$ and hence $z_q=z_s=z_K=1$.
The fit  in this case is very poor, $\chi^2/$d.o.f.=31.2/6.
The quality of the fit gets worse when $\phi_{\eta G}$ and $\phi_{\eta^\prime G}$ are set to zero,
$\chi^2/$d.o.f.=45.9/8 with $\phi_P=(41.1\pm 1.1)^\circ$.

Clearly, in order to obtain a good fit one has to relax the constraint imposed on the overlapping parameters.
Hence, we begin to discuss the second of the possibilities, that is to say,
to leave the $z$'s free and restrict the gluon content of the $\eta$ or $\eta^\prime$ meson.
However, as a matter of comparison, we first consider the absence of gluonium in both mesons,
\emph{i.e.~}$\phi_{\eta G}=\phi_{\eta^\prime G}=0$.
In addition, we also fix the vector mixing angle $\phi_V$ to its measured value
$\tan\phi_V=+0.059\pm 0.004$ or $\phi_V=(3.4\pm 0.2)^\circ$ \cite{Dolinsky:1991vq}
and the ratio of constituent quark masses to $\bar m/m_s\simeq 1/1.45$.
The fit in this case is not yet satisfactory, $\chi^2/$d.o.f.=14.0/7.
The quality of the fit improves when the ratio $\bar m/m_s$ is left free, 
$\chi^2/$d.o.f.=7.6/6 with $m_s/\bar m=1.24\pm 0.07$.
If $\phi_V$ is also left free, the final result of the fit gives $\chi^2/$d.o.f.=4.4/5 with
\begin{equation}
\label{zphiP}
\begin{array}{c}
g=0.72\pm 0.01\ \mbox{GeV$^{-1}$}\ ,\quad \phi_P=(41.5\pm 1.2)^\circ\ ,\quad
\phi_V=(3.2\pm 0.1)^\circ\ ,\\[2ex]
\frac{m_s}{\bar m}=1.24\pm 0.07\ ,\quad
z_q=0.86\pm 0.03\ ,\quad z_s=0.78\pm 0.05\ ,\quad z_K=0.89\pm 0.03\ .
\end{array}
\end{equation}
The fitted values for the two mixing angles $\phi_P$ and $\phi_V$ are in good agreement with most results coming from other analyses using complementary information
(see, for instance, Ref.~\cite{Bramon:1997va} and references therein).
Our value for the pseudoscalar mixing angle also agrees with the latest measurement from KLOE,
$\phi_P=(41.4\pm 1.0)^\circ$ \cite{Ambrosino:2006gk}.
The free parameters $z$'s are specific of our approach and are not fixed to one as in
previous analyses \cite{Isgur:1976hg,O'Donnell:1981sj}.
As mentioned, if we fix the $z$'s to unity, the fit gets much worse ($\chi^2/$d.o.f.=45.9/8).
This shows that allowing for different overlaps of quark-antiquark wave functions and, in particular, for those coming from the gluon anomaly affecting only the $\eta$ and $\eta^\prime$ singlet component, is indeed relevant.

Finally, the symmetry breaking parameter $m_s/\bar m$ which is essential to adjust the ratio between the two $K^\ast$-$K$ transitions is affected by large uncertainties due to the difficulty in extracting the neutral and charged $K^\ast\to K\gamma$ widths from
Primakoff-effect analyses\footnote{
The neutral and charged $K^\ast\to K\gamma$ transitions have been measured only by one and two experimental groups, respectively, and seem to need further confirmations \cite{Yao:2006px}.}.
For these reasons, we have performed a new fit ignoring the two $K^\ast\to K\gamma$ channels.
This new fit  obviously gives the same results for $g$, $\phi_{P,V}$ and $z_q$ whereas
$m_s/\bar m$ and $z_s$ always appear in the combination $z_s\bar m/m_s$
which is fitted to $0.63\pm 0.02$, in agreement with our previous values in Eq.~(\ref{zphiP}).
Therefore, our results are insensitive to eventual modifications of future and desirable new data on
$K^\ast\to K\gamma$ transitions.

\TABLE[t]{
\centering
\begin{tabular}{cccc}
\hline\\[-1.5ex]
Transition & $g_{VP\gamma}^{\rm exp}$(PDG) & 
$g_{VP\gamma}^{\rm th}$(Fit 1) & $g_{VP\gamma}^{\rm th}$(Fit 2)\\[1ex]
\hline\\[-1.5ex]
$\rho^0\rightarrow\eta\gamma$ &
$0.475\pm 0.024$ & $0.461\pm 0.019$ & $0.464\pm 0.030$\\[1ex]
$\eta^\prime\rightarrow\rho^0\gamma$ &
$0.41\pm 0.03$ & $0.41\pm 0.02$ & $0.40\pm 0.04$\\[1ex]
$\omega\rightarrow\eta\gamma$ &
$0.140\pm 0.007$ & $0.142\pm 0.007$ & $0.143\pm 0.010$\\[1ex]
$\eta^\prime\rightarrow\omega\gamma$ &
$0.139\pm 0.015$ & $0.149\pm 0.006$ & $0.146\pm 0.014$\\[1ex]
$\phi\rightarrow\eta\gamma$ &
$0.209\pm 0.002$ & $0.209\pm 0.018$ & $0.209\pm 0.013$\\[1ex]
$\phi\rightarrow\eta^\prime\gamma$ &
$0.22\pm 0.01$ & $0.22\pm 0.02$ & $0.22\pm 0.02$\\[1ex]
\hline
\end{tabular}
\caption{Comparison between the experimental values $g_{VP\gamma}^{\rm exp}$ (in GeV$^{-1}$)
for the various $(V,P)\rightarrow (P,V)\gamma$ transitions, with $P=\eta,\eta^\prime$,
taken from the PDG \cite{Yao:2006px}
and the corresponding predictions for $g_{VP\gamma}^{\rm th}$
from Eqs.~(\ref{zphiP}) ---Fit 1--- and (\ref{zphiPphietapG}) ---Fit 2---.}
\label{table1}
}
In Table \ref{table1}, we present a comparison between experimental data for the relevant
$VP\gamma$ transitions with $P=\eta,\eta^\prime$ and the corresponding theoretical predictions
(in absolute value) calculated from the fitted values in Eq.~(\ref{zphiP}).
We do not include in that comparison neither $V\to\pi\gamma$ nor $K^\ast\to K\gamma$ modes since they are used to constrain the complementary parameters $g, \phi_V, m_s/\bar m$ and $z_K$.
The agreement is very good and all the predictions coincide with the experimental values within
$1\sigma$.
Notice the small experimental error for the $g_{\phi\eta\gamma}$ coupling as compared to the theoretical one, which, as we will see in Sec.~\ref{approaches}, serves to highly constrain the allowed values for the $\eta$-$\eta^\prime$ mixing angle $\phi_P$.

Now that we have performed a fit under the hypothesis of no gluonium we return to the main issue of this analysis, the phenomenological determination of the gluon content of the $\eta$ and $\eta^\prime$ mesons.
As stated before, a simultaneous fit of the overlapping parameters $z_{q,s}$ and the three mixing angles
$\phi_P$, $\phi_{\eta G}$ and $\phi_{\eta^\prime G}$ is not feasible.
Therefore, we first assume $\phi_{\eta G}=0$, \textit{i.e.}~$Z_\eta=0$,
and then proceed to fit the gluonic content of the $\eta^\prime$ wave function under this assumption.
The results of the new fit are\footnote{
There is a sign ambiguity in $\phi_{\eta^\prime G}$ that cannot be decided since this angle
enters into $X_{\eta^\prime}$ and $Y_{\eta^\prime}$ through a cosine.}
\begin{equation}
\label{zphiPphietapG}
\begin{array}{c}
g=0.72\pm 0.01\ \mbox{GeV$^{-1}$}\ ,\quad \frac{m_s}{\bar m}=1.24\pm 0.07\ ,\quad
\phi_V=(3.2\pm 0.1)^\circ\ ,\\[2ex]
\phi_P=(41.4\pm 1.3)^\circ\ ,\quad |\phi_{\eta^\prime G}|=(12\pm 13)^\circ\ ,\\[2ex]
z_q=0.86\pm 0.03\ ,\quad z_s=0.79\pm 0.05\ ,\quad z_K=0.89\pm 0.03\ ,
\end{array}
\end{equation}
with $\chi^2/$d.o.f.=4.2/4.
The quality of the fit is similar to the one obtained assuming a vanishing gluonic admixture for both mesons ($\chi^2/$d.o.f.=4.4/5).
The fitted values for $z_q$ and $z_s$ are compatible with those of Eq.~(\ref{zphiP}).
The result obtained for $\phi_{\eta^\prime G}$ suggests a very small amount of gluonium in the
$\eta^\prime$ wave function, in fact compatible with zero within $1\sigma$.
Using Eq.~(\ref{gluoniumetap}) to calculate $Z_{\eta^\prime}$ from $\phi_{\eta^\prime G}$ gives
$|Z_{\eta^\prime}|=0.2\pm 0.2$.
This is one of the main results of our analysis.
Accepting the absence of gluonium for the $\eta$ meson, the gluonic content of the $\eta^\prime$ wave function amounts to $|\phi_{\eta^\prime G}|=(12\pm 13)^\circ$ or $Z_{\eta^\prime}^2=0.04\pm 0.09$.

In other words, our values for $\phi_P$ and $\phi_{\eta^\prime G}$ (or $Z_{\eta^\prime}$)
contrast with those reported by KLOE recently, $\phi_P=(39.7\pm 0.7)^\circ$ and
$|\phi_{\eta^\prime G}|=(22\pm 3)^\circ$ ---or $Z_{\eta^\prime}^2=0.14\pm 0.04$---
\cite{Ambrosino:2006gk}.
As indicated in Sec.~\ref{intro}, a possible explanation of this discrepancy could be the use in
Ref.~\cite{Ambrosino:2006gk} of old values for the overlapping parameters that the present analysis tries to update.
In Table \ref{table1}, we also include the theoretical predictions for the various transitions involving
$\eta$ or $\eta^\prime$ calculated from the fitted values in Eq.~(\ref{zphiPphietapG}).
As expected, there is no significant difference between the values obtained allowing for gluonium (Fit 2) or not (Fit 1) in the $\eta^\prime$ wave function.
Likewise, we predict the value of the ratio
\begin{equation}
\label{Rphi}
R_\phi\equiv\frac{\Gamma(\phi\to\eta^\prime\gamma)}{\Gamma(\phi\to\eta\gamma)}=
\cot^2\phi_P\cos^2\phi_{\eta^\prime G}
\left(1-\frac{m_s}{\bar m}\frac{z_q}{z_s}\frac{\tan\phi_V}{\sin 2\phi_P}\right)^2
\left(\frac{p_{\eta^\prime}}{p_\eta}\right)^3\ ,
\end{equation}
to be $(4.7\pm 0.6)\times 10^{-3}$, in agreement with the experimental value in
Ref.~\cite{Yao:2006px}, $(4.8\pm 0.5)\times 10^{-3}$,
and the most recent measurement by KLOE \cite{Ambrosino:2006gk},
$(4.77\pm 0.09_{\rm stat}\pm 0.19_{\rm syst})\times 10^{-3}$.

For completeness, we perform another fit assuming from the beginning a null gluonic content for the
$\eta^\prime$ meson.
Consequently, we fix $\phi_{\eta^\prime G}=0$ and leave $\phi_{\eta G}$ free.
The results obtained are the following:
\begin{equation}
\label{zphiPphietaG}
\begin{array}{c}
g=0.72\pm 0.01\ \mbox{GeV$^{-1}$}\ ,\quad \frac{m_s}{\bar m}=1.24\pm 0.07\ ,\quad
\phi_V=(3.2\pm 0.1)^\circ\ ,\\[2ex]
\phi_P=(41.5\pm 1.3)^\circ\ ,\quad |\phi_{\eta G}|\simeq 0^\circ\ ,\\[2ex]
z_q=0.86\pm 0.04\ ,\quad z_s=0.78\pm 0.06\ ,\quad z_K=0.89\pm 0.03\ ,
\end{array}
\end{equation}
with $\chi^2/$d.o.f.=4.4/4.
The fitted value for $\phi_{\eta G}$ is very close to zero.
For that reason, it is better to express this value in terms of the more common $Z_\eta$ parameter.
As a result, one gets $Z_\eta^2=0.00\pm 0.12$, thus showing a vanishing gluonium contribution in the
$\eta$ wave function.
This is a second important result of our analysis which complements the one discussed after
Eq.~(\ref{zphiPphietapG}).
To sum up, the current experimental data on $VP\gamma$ transitions seem to indicate within our model a negligible gluonic content for the $\eta$ and $\eta^\prime$ mesons.

\TABLE[t]{
\centering
\begin{tabular}{cccc}
\hline\\[-1.5ex]
Transition & $g_{VP\gamma}^{\rm exp}$(latest) & 
$g_{VP\gamma}^{\rm th}$(Fit 3) & $g_{VP\gamma}^{\rm th}$(Fit 4)\\[1ex]
\hline\\[-1.5ex]
$\rho^0\rightarrow\eta\gamma$ &
$0.429\pm 0.023$ & $0.436\pm 0.017$ & $0.437\pm 0.028$\\[1ex]
$\eta^\prime\rightarrow\rho^0\gamma$ &
$0.41\pm 0.03$ (PDG) & $0.40\pm 0.02$ & $0.40\pm 0.04$\\[1ex]
$\omega\rightarrow\eta\gamma$ &
$0.136\pm 0.007$ & $0.134\pm 0.006$ & $0.134\pm 0.009$\\[1ex]
$\eta^\prime\rightarrow\omega\gamma$ &
$0.139\pm 0.015$ (PDG) & $0.146\pm 0.006$ & $0.146\pm 0.013$\\[1ex]
$\phi\rightarrow\eta\gamma$ &
$0.214\pm 0.003$ & $0.214\pm 0.017$ & $0.214\pm 0.012$\\[1ex]
$\phi\rightarrow\eta^\prime\gamma$ &
$0.216\pm 0.005$ & $0.216\pm 0.019$ & $0.216\pm 0.018$\\[1ex]
\hline
\end{tabular}
\caption{The same as in Table \ref{table1} but for $g_{VP\gamma}^{\rm th}$
from Eqs.~(\ref{zphiPlatest}) ---Fit 3--- and (\ref{zphiPphietapGlatest}) ---Fit 4---  compared to the latest $g_{VP\gamma}^{\rm exp}$ (in GeV$^{-1}$) from Refs.~\cite{Ambrosino:2006gk,Achasov:2006dv}.}
\label{table2}
}
A final exercise we have done is to check whether the very recent measurements
(not included in Ref.~\cite{Yao:2006px}) on
$\rho,\omega,\phi\to\eta\gamma$ from the SND Coll.~\cite{Achasov:2006dv} and
$\phi\to\eta^\prime\gamma$ from KLOE \cite{Ambrosino:2006gk}
modify the results of our analysis.
The values of the couplings associated to these new data are displayed in Table \ref{table2}.
As shown, the central values are nearly the same as those from Ref.~\cite{Yao:2006px}, except for
$\rho\to\eta\gamma$, whereas the errors for $\rho,\omega,\phi\to\eta\gamma$ are comparable to the world averages and the error for $\phi\to\eta^\prime\gamma$ is reduced by a factor of three.
Assuming absence of gluonium in the $\eta$ and $\eta^\prime$ wave functions
the results of the fit are
\begin{equation}
\label{zphiPlatest}
\phi_P=(42.7\pm 0.7)^\circ\ ,\quad z_q=0.83\pm 0.03\ ,\quad z_s=0.79\pm 0.05\ ,
\end{equation}
with $\chi^2/$d.o.f.=4.0/5.
Accepting now a gluonic admixture only in the $\eta^\prime$, one obtains
\begin{equation}
\label{zphiPphietapGlatest}
\phi_P=(42.6\pm 1.1)^\circ\ ,\quad |\phi_{\eta^\prime G}|=(5\pm 21)^\circ\ ,
\quad z_q=0.83\pm 0.03\ ,\quad z_s=0.79\pm 0.05\ ,
\end{equation}
with $\chi^2/$d.o.f.=4.0/4.
The values of the remaining parameters are the same as in Eqs.~(\ref{zphiP}) and (\ref{zphiPphietapG}), respectively.
As one can see, the central values for the mixing angle $\phi_P$ slightly increase while the ones for
$\phi_{\eta^\prime G}$ (or $Z_{\eta^\prime}^2=0.01\pm 0.07$) and $z_q$ decrease.
However, in both cases the quality of the fit is as good as it was when only the experimental data provided by Ref.~\cite{Yao:2006px} were considered.
Moreover, the fitted values of all the parameters are compatible within $1\sigma$ with those of
Eqs.~(\ref{zphiP},\ref{zphiPphietapG}).
The corresponding theoretical predictions for the $g_{VP\gamma}$ couplings with $P=\eta,\eta^\prime$
calculated from the fitted values in Eqs.~(\ref{zphiPlatest}) ---Fit 3--- and 
(\ref{zphiPphietapGlatest}) ---Fit 4--- are also shown in Table \ref{table2}.
Note in this case the small experimental error for $g_{\phi\eta^\prime\gamma}$ which will constrain even more the allowed values for $Z_{\eta^\prime}^2$ (see Sec.~\ref{approaches}).
In conclusion, the latest experimental data seem to confirm the null gluonic content of the
$\eta$ and $\eta^\prime$ wave functions.

\section{Comparison with other approaches}
\label{approaches}
Our main results can also be displayed graphically following
Refs.~\cite{Ambrosino:2006gk,Rosner:1982ey,Kou:1999tt}.
This will serve us to present the bounds obtained using our approach and compare them with other approaches.
Very briefly, the analysis of Ref.~\cite{Rosner:1982ey} is based on the $SU(3)$ quark model supplemented by the $SU(3)$ breaking parameter $\bar m/m_s$ dealing with the difference of the down and strange quark magnetic moments but without taking into account the vector mixing angle $\phi_V$ or the different vector-pseudoscalar overlapping parameters, both specific features of our approach.
The experimental analysis of Ref.~\cite{Ambrosino:2006gk} is based on the same approach including the latter two features.
Finally, the analysis of Ref.~\cite{Kou:1999tt} utilizes the $SU(3)$ quark model and the vector meson dominance model (VMD).
In this framework, the transition amplitudes are expressed in terms of masses and decay constants affected by $SU(3)$ breaking effects.

\FIGURE[t]{
\centerline{\includegraphics[width=0.75\textwidth]{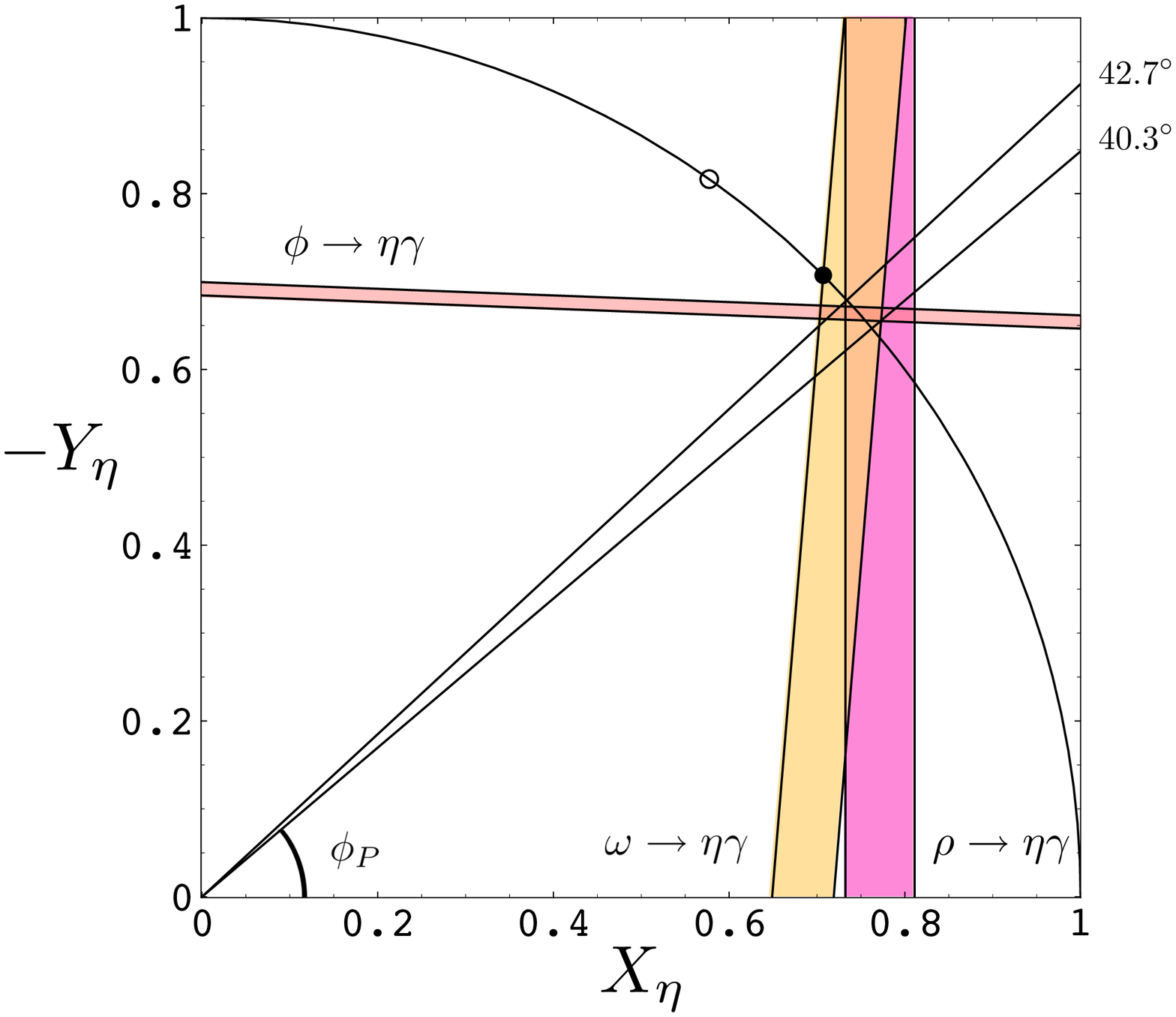}}  
\caption{Constraints on non-strange ($X_\eta$) and strange ($Y_\eta$) quarkonium mixing coefficients in the $\eta$ as defined in Eq.~(\ref{mixingP}).
The circular boundary denotes the constraint $X_\eta^2+Y_\eta^2\leq 1$.
The mixing solutions corresponding to the $\eta$ being a pure octet
($X_\eta=-\frac{1}{\sqrt{2}}Y_\eta=\frac{1}{\sqrt{3}}$) ---open circle--- and
($X_\eta=Y_\eta=\frac{1}{\sqrt{2}}$) ---closed circle--- are shown.
The straight lines for the upper and lower bounds of the mixing solution in
\emph{absence of gluonium}, $\phi_P=(41.5\pm 1.2)^\circ$, are also shown.
The vertical and inclined bands are the regions for $X_\eta$ and $Y_\eta$ allowed by the experimental couplings of the $(\rho,\omega,\phi)\to\eta\gamma$ transitions in Table \ref{table1}.}
\label{ploteta}
}
In Fig.~\ref{ploteta}, we plot the regions for the $X_\eta$ and $Y_\eta$ parameters which are allowed by the experimental couplings of the
$\rho\to\eta\gamma$, $\omega\to\eta\gamma$ and $\phi\to\eta\gamma$ transitions
(see Table \ref{table1}).
The limits of the bands are given at 68\% CL or $1\sigma$.
For the rest of the parameters involved in the determination of the allowed regions we have used the fitted values for $g$, $\phi_V$, $m_s/\bar m$, $z_q$, $z_s$, and $z_K$ from Eq.~(\ref{zphiPphietaG}).
In addition to the bands, we have also plotted the circular boundary denoting the constraint
$X_\eta^2+Y_\eta^2\leq 1$ as well as the favoured region for the $\eta$-$\eta^\prime$ mixing angle
assuming the \emph{absence of gluonium}, $40.3^\circ\leq\phi_P\leq 42.7^\circ$, obtained at $1\sigma$ from the corresponding fitted value in Eq.~(\ref{zphiP}).
As one can see, there exists a perfect intersection region of the three bands located precisely on the circumference at the region preferred by the mixing angle, thus indicating a vanishing gluonium contribution in the $\eta$ wave function, \emph{i.e.}~$|Z_\eta|=0$.
The small size of the intersection region, mainly due to the small experimental error in the
$g_{\phi\eta\gamma}$ coupling, and its precise location on the circumference gives strong evidence in favour of the former statement.
Notice also the importance of considering the vector mixing angle $\phi_V$ different from zero, which translates into a finite slope for the $(\omega,\phi)\to\eta\gamma$ bands, in the result obtained.
This contrasts with the analysis of Refs.~\cite{Rosner:1982ey,Kou:1999tt} where the vector mixing angle was not taken into account at that time because their effects were covered by the bigger experimental errors of the bands.
With the present available experimental data, fixing the value of this angle to zero would have lead to an incompatible solution for the three bands.
For completeness, the points corresponding to the $\eta$ being a pure octet
($X_\eta=-\frac{1}{\sqrt{2}}Y_\eta=\frac{1}{\sqrt{3}}$) and the ``democratic'' mixing solution
($X_\eta=Y_\eta=\frac{1}{\sqrt{2}}$) are also shown in Fig.~\ref{ploteta} as an open and closed circle, respectively.
The latter solution, which is now excluded, was still acceptable in the analysis of
Ref.~\cite{Rosner:1982ey}, where the upper bound $|Z_\eta|\lesssim 0.4$ was obtained from the processes $(\rho,\phi)\to\eta\gamma$ and $\eta\to\gamma\gamma$.
In Ref.~\cite{Kou:1999tt}, after correcting some inconsistent data, there was no region where all three constraints, coming from the $(\omega,\phi)\to\eta\gamma$ and $\eta\to\gamma\gamma$ decays, overlap.

\FIGURE[t]{
\centerline{\includegraphics[width=0.70\textwidth]{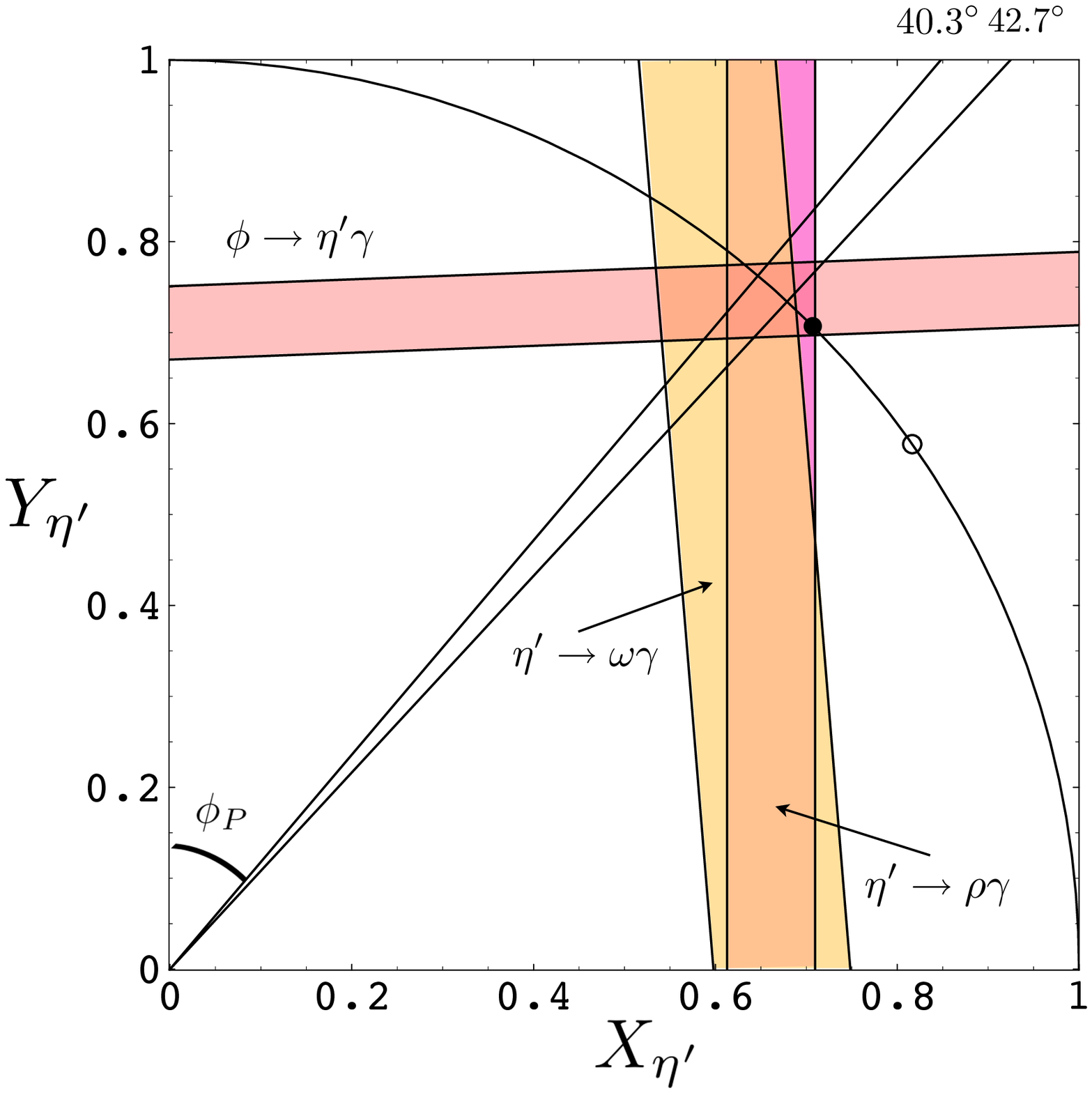}}  
\caption{Constraints on non-strange ($X_{\eta^\prime}$) and strange ($Y_{\eta^\prime}$) quarkonium mixing coefficients in the $\eta^\prime$.
The mixing solutions corresponding to the $\eta^\prime$ being a pure singlet
($X_{\eta^\prime}=\sqrt{2}Y_{\eta^\prime}=\frac{1}{\sqrt{3}}$) ---open circle--- and
($X_{\eta^\prime}=Y_{\eta^\prime}=\frac{1}{\sqrt{2}}$) ---closed circle--- are shown.
The vertical and inclined bands are the regions for $X_{\eta^\prime}$ and $Y_{\eta^\prime}$ allowed by the experimental couplings of the $\eta^\prime\to(\rho,\omega)\gamma$ and
$\phi\to\eta^\prime\gamma$ transitions.}
\label{plotetap}
}
The situation is not so constrained for the $\eta^\prime$ as we proceed to discuss.
In Fig.~\ref{plotetap}, we plot the regions for the $X_{\eta^\prime}$ and $Y_{\eta^\prime}$ parameters which are allowed by the experimental couplings of the
$\eta^\prime\to\rho\gamma$, $\eta^\prime\to\omega\gamma$ and $\phi\to\eta^\prime\gamma$ transitions.
The remaining parameters are taken from Eq.~(\ref{zphiPphietapG}).
Again, we also plot the constraint implied by $X_{\eta^\prime}^2+Y_{\eta^\prime}^2\leq 1$ and the favoured region for the $\eta$-$\eta^\prime$ mixing angle in absence of gluonium.
This time there exists an intersection region of the three bands inside and on the circumference.
As most of this region is interior but close to the circular boundary it may well indicate a small but non necessarily zero gluonic content of the $\eta^\prime$.
Indeed, we have found $Z_{\eta^\prime}^2=0.04\pm 0.09$ (or $|Z_{\eta^\prime}|=0.2\pm 0.2$) or using the angular description $|\phi_{\eta^\prime G}|=(12\pm 13)^\circ$.
The size of the error is precisely what prevent us from drawing a definite conclusion concerning the amount of gluonium in the $\eta^\prime$ wave function.
More refined experimental data, particularly for the $\phi\to\eta^\prime\gamma$ channel, will contribute decisively to clarify this issue (see below).
Clearly, the inclusion of this process is of major importance for the determination of the gluonic admixture in the $\eta^\prime$, as observed for the first time in Ref.~\cite{Rosner:1982ey}.
In this latter analysis, where $\eta^\prime\to\rho\gamma$ and $\eta^\prime\to\gamma\gamma$ were used in addition to different values for $\phi\to\eta^\prime\gamma$, the absence of a significant constraint on $Y_{\eta^\prime}$ was keenly felt.
However, the mixing solution $X_{\eta^\prime}=Y_{\eta^\prime}=\frac{1}{\sqrt{2}}$
---closed circle in Fig.~\ref{plotetap}--- was still acceptable.
It was not the case for the solution identifying the $\eta^\prime$ as a pure singlet,
$X_{\eta^\prime}=\sqrt{2}Y_{\eta^\prime}=\frac{1}{\sqrt{3}}$, ---open circle in Fig.~\ref{plotetap}---.
In the present analysis, the ``democratic'' mixing solution is excluded at the $1\sigma$ level whereas the singlet solution is clearly excluded.
In Ref.~\cite{Kou:1999tt}, where $\eta^\prime\to\omega\gamma$ was also included in the analysis, the maximum gluonic admixture in the $\eta^\prime$ was obtained to be 26\% for $\theta_P=-11^\circ$
(or $\phi_P\simeq 44^\circ$).
In our case, for $Z_{\eta^\prime}^2=0.04\pm 0.09$, one gets
$R=\frac{Z_{\eta^\prime}}{X_{\eta^\prime}+Y_{\eta^\prime}+Z_{\eta^\prime}}=(13\pm 13)\%$
for $\phi_P=41.4^\circ$ (or $\theta_P=-13.4^\circ$).
The same analysis also anticipated that the existence of a gluonic content for the $\eta^\prime$ would be excluded for large $|\theta_P|$ (or small $\phi_P$).
Finally, the KLOE analysis in Ref.~\cite{Ambrosino:2006gk} has found a solution allowing for gluonium with $\phi_P=(39.7\pm 0.7)^\circ$ and $Z_{\eta^\prime}^2=0.14\pm 0.04$
---or $|\phi_{\eta^\prime G}|=(22\pm 3)^\circ$---, in disagreement, particularly for the $Z_{\eta^\prime}^2$ value, with our results.
A second analysis without the constraint from the $\eta^\prime\to\rho\gamma$ decay, which is the less precise, gives $\phi_P=(39.8\pm 0.8)^\circ$ and $Z_{\eta^\prime}^2=0.13\pm 0.04$, thus not modifying their conclusions.
\FIGURE[t]{
\centerline{\includegraphics[width=0.45\textwidth,angle=90]{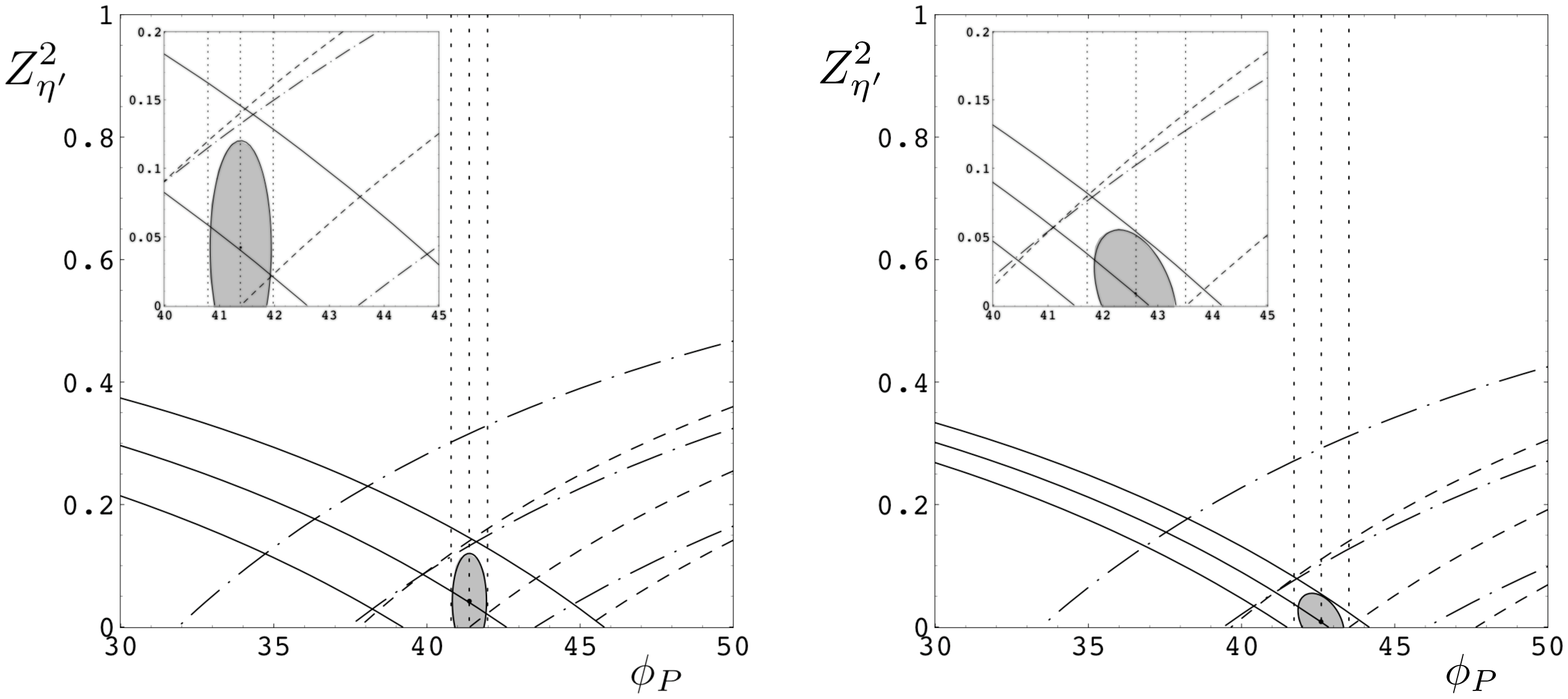}}  
\caption{The gray ellipse in the $(\phi_P,Z_{\eta^\prime}^2)$ plane corresponds to the allowed region at 68\% CL of the solution in Eq.~(\ref{zphiPphietapG}) ---\emph{left plot}--- and
Eq.~(\ref{zphiPphietapGlatest}) ---\emph{right plot}---, respectively, assuming the presence of gluonium.
The different bands are the regions for $\phi_P$ and $Z_{\eta^\prime}^2$ allowed by the experimental couplings of the $\eta^\prime\to\rho\gamma$ (dashed line),
$\eta^\prime\to\omega\gamma$ (dot-dashed line), $\phi\to\eta\gamma$ (dotted line), and
$\phi\to\eta^\prime\gamma$ (solid line) transitions in Table \ref{table1} ---\emph{left plot}---
and Table \ref{table2} ---\emph{right plot}---, respectively.}
\label{italians}
}
\noindent
To make this difference more graphical, we follow Ref.~\cite{Ambrosino:2006gk} and
plot in Fig.~\ref{italians} ---\emph{left plot}--- the constraints from $\eta^\prime\to(\rho,\omega)\gamma$ and $\phi\to(\eta,\eta^\prime)\gamma$ in the $(\phi_P,Z_{\eta^\prime}^2)$ plane together with the
68\% CL allowed region for gluonium as obtained from Eq.~(\ref{zphiPphietapG}).
The point corresponding to the preferred solution, $(\phi_P,Z_{\eta^\prime}^2)=(41.4^\circ,0.04)$,
is also shown.
The allowed region is very constrained in the $\phi_P$ axis by the experimental value of the
$g_{\phi\eta\gamma}$ coupling, whose vertical band denotes its non dependence on
$Z_{\eta^\prime}^2$.
The other three bands, all dependent on $\phi_P$ and $Z_{\eta^\prime}^2$, constrain the amount of gluonium down to a value compatible with zero within $1\sigma$.
As mentioned before, our result differs from the KLOE's one, where
$(\phi_P,Z_{\eta^\prime}^2)=(39.7^\circ,0.14)$ is the preferred solution and the allowed region for gluonium is far from the $\phi_P$ axis
(see Fig.~5 in Ref.~\cite{Ambrosino:2006gk}).
The value compatible with zero we have obtained for the gluonic content of the $\eta^\prime$ wave function is fully confirmed from a graphical point of view as soon as one includes the latest measurements of the $\phi\to(\eta,\eta^\prime)\gamma$ decays from
Refs.~\cite{Ambrosino:2006gk,Achasov:2006dv}.
As seen in Fig.~\ref{italians} ---\emph{right plot}---, the allowed region for gluonium decreases due to the smaller experimental error of the $g_{\phi\eta^\prime\gamma}$ coupling (although the error of
$g_{\phi\eta\gamma}$ increases a little) and its central point (the preferred solution) is now located at
$(\phi_P,Z_{\eta^\prime}^2)=(42.6^\circ,0.01)$, even much closer to the $\phi_P$ axis than before.

In all the other approaches discussed so far, the decays $(\eta,\eta^\prime)\to\gamma\gamma$ were also included in the analyses.
As shortly stated in Sec.~\ref{intro}, in our approach we do not take into account these processes since they are well explained using a two mixing angle scenario
\cite{Escribano:2005qq,Feldmann:1999uf}.
Nevertheless, our framework can be extended to describe $P^0\to\gamma\gamma$ decays in a similar way to $V\to P\gamma$ and $P\to V\gamma$.
In such case, the needed ratios of decay widths are found to be
\begin{equation}
\label{P0gg}
\begin{array}{l}
\frac{\Gamma(\eta\to\gamma\gamma)}{\Gamma(\pi^0\to\gamma\gamma)}=
\frac{1}{9}\left(\frac{m_\eta}{m_{\pi^0}}\right)^3
\left(5\,\tilde z_q\,X_\eta+\sqrt{2}\frac{\bar m}{m_s}\tilde z_s\,Y_\eta\right)^2\ ,\\[2ex]
\frac{\Gamma(\eta^\prime\to\gamma\gamma)}{\Gamma(\pi^0\to\gamma\gamma)}=
\frac{1}{9}\left(\frac{m_{\eta^\prime}}{m_{\pi^0}}\right)^3
\left(5\,\tilde z_q\,X_{\eta^\prime}+\sqrt{2}\frac{\bar m}{m_s}\tilde z_s\,Y_{\eta^\prime}\right)^2\ ,
\end{array}
\end{equation}
where, in addition to the parameter $\bar m/m_s$ related to the different quark spin-flip effects of the emission of one photon, we have to introduce two new $SU(3)$ breaking parameters,
$\tilde z_q\equiv\tilde C_q/\tilde C_\pi$ and $\tilde z_s\equiv\tilde C_s/\tilde C_\pi$, dealing with the spatial wave functions effects of quark-antiquark annihilation into the other photon.
These new parameters are analogous to the overlapping parameters, $z_q$ and $z_s$, of the
$VP\gamma$ transitions, and for the same reason, the $U(1)_A$ anomaly, we also distinguish between the quark-antiquark annihilation effects in $|\pi\rangle$ and $|\eta_q\rangle$.
For comparison, in Ref.~\cite{Rosner:1982ey} the former $SU(3)$ breaking effects are not considered, \emph{i.e.~}$\tilde z_q=\tilde z_s=1$ and $\bar m/m_s=1$, while in Ref.~\cite{Kou:1999tt}, where the
$P^0\to\gamma\gamma$ decays are characterized by means of pseudoscalar decay constants, one identifies $\frac{f_q}{f_\pi}={\tilde z_q}^{-1}$ and $\frac{f_s}{f_\pi}=\frac{m_s}{\bar m}{\tilde z_s}^{-1}$.
As seen in Eq.~(\ref{P0gg}), these ratios are not useful for fixing, within our approach, the mixing parameters of the $\eta$-$\eta^\prime$ system since they also depend on the unknown values of
$\tilde z_q$ and $\tilde z_s$.
However, in order to check the consistency of this extended framework, one could use for the mixing parameters and $\bar m/m_s$ the results obtained from the analysis of $V\to P\gamma$ and
$P\to V\gamma$ decays and then fix the values of $\tilde z_{q,s}$ from the two ratios under consideration.
In this case, one gets $\tilde z_q=0.96\pm 0.03$ and $\tilde z_s=0.79\pm 0.13$ and hence
$f_q=(1.05\pm 0.03)f_\pi$ and $f_s=(1.57\pm 0.28)f_\pi$, in agreement with the values
$f_q=(1.10\pm 0.03)f_\pi$ and $f_s=(1.66\pm 0.07)f_\pi$ found in Ref.~\cite{Escribano:2005qq},
thus showing the consistency of the mixing parameters obtained and of the whole approach.

\section{Summary and conclusions}
\label{conclusions}
In this work we have performed a phenomenological analysis of radiative $V\to P\gamma$ and
$P\to V\gamma$ decays with the purpose of determining the gluon content of the $\eta$ and
$\eta^\prime$ mesons.
The present approach is based on a conventional $SU(3)$ quark model supplemented with two sources of $SU(3)$ breaking, the use of constituent quark masses with $m_s>\bar m$, thus making the
$VP\gamma$ magnetic dipole transitions to distinguish between photon emission from strange or
non-strange quarks, and the different spatial extensions for each $P$ or $V$ isomultiplet which induce different overlaps between the $P$ and $V$ wave functions.
The use of these different overlapping parameters ---a specific feature of our analysis--- is shown to be of primary importance in order to reach a good agreement.

Our conclusions are the following.
First, the current experimental data on $VP\gamma$ transitions indicate within our model a negligible gluonic content for the $\eta$ and $\eta^\prime$ mesons, $Z_\eta^2=0.00\pm 0.12$ and
$Z_{\eta^\prime}^2=0.04\pm 0.09$.
Second, accepting the absence of gluonium for the $\eta$ meson, the gluonic content of the 
$\eta^\prime$ wave function amounts to $|\phi_{\eta^\prime G}|=(12\pm 13)^\circ$ and the
$\eta$-$\eta^\prime$ mixing angle is found to be $\phi_P=(41.4\pm 1.3)^\circ$.
Third, imposing the absence of gluonium for both mesons one finds $\phi_P=(41.5\pm 1.2)^\circ$,
in agreement with the former result.
Fourth, the latest experimental data on $(\rho,\omega,\phi)\to\eta\gamma$ and
$\phi\to\eta^\prime\gamma$ decays confirm the null gluonic content of the $\eta$ and
$\eta^\prime$ wave functions.
Finally, we would like to stress that more refined experimental data, particularly for the
$\phi\to\eta^\prime\gamma$ channel, will contribute decisively to clarify this issue.

\section*{Acknowledgements}
R.E. thanks A.~Bramon for a critical reading of the manuscript.
This work was supported in part by the Ramon y Cajal program (R.E.),
the Ministerio de Educación y Ciencia under grant FPA2005-02211,
the EU Contract No.~MRTN-CT-2006-035482, ``FLAVIAnet'', and
the Generalitat de Catalunya under grant 2005-SGR-00994.

\appendix
\section{Euler angles}
\label{decomposition}
In presence of gluonium, the wave functions of the $\eta$ and $\eta^\prime$ mesons and the 
glueball-like state $\iota$ can be decomposed as
\begin{equation}
\label{wavefunctions}
\begin{array}{rcl}
|\eta\rangle &=& X_\eta|\eta_q\rangle+Y_\eta|\eta_s\rangle+Z_\eta|G\rangle\ ,\\[1ex]
|\eta^\prime\rangle &=&
X_{\eta^\prime}|\eta_q\rangle+Y_{\eta^\prime}|\eta_s\rangle+Z_{\eta^\prime}|G\rangle\ ,\\[1ex]
|\iota\rangle &=& X_\iota|\eta_q\rangle+Y_\iota|\eta_s\rangle+Z_\iota|G\rangle\ ,
\end{array}
\end{equation}
where $|\eta_q\rangle\equiv\frac{1}{\sqrt{2}}|u\bar u+d\bar d\rangle$, $|\eta_s\rangle=|s\bar s\rangle$
and $|G\rangle\equiv|\mbox{gluonium}\rangle$.
The $\iota$ or $\eta(1440)$ state, which we refrain from discussing here, could be identified with the
$\eta(1405)$ pseudoscalar resonance (see Ref.~\cite{Yao:2006px} for details).
The nine coefficients in Eq.~(\ref{wavefunctions}) are constrained by three normalization conditions
\begin{equation}
\label{normalization}
\begin{array}{r}
X_\eta^2+Y_\eta^2+Z_\eta^2=1\ ,\\[1ex]
X_{\eta^\prime}^2+Y_{\eta^\prime}^2+Z_{\eta^\prime}^2=1\ ,\\[1ex]
X_\iota^2+Y_\iota^2+Z_\iota^2=1\ ,
\end{array}
\end{equation}
and three orthogonality conditions
\begin{equation}
\label{orthogonality}
\begin{array}{r}
X_\eta X_{\eta^\prime}+Y_\eta Y_{\eta^\prime}+Z_\eta Z_{\eta^\prime}=0\ ,\\[1ex]
X_\eta X_\iota+Y_\eta Y_\iota+Z_\eta Z_\iota=0\ ,\\[1ex]
X_{\eta^\prime}X_\iota+Y_{\eta^\prime}Y_\iota+Z_{\eta^\prime}Z_\iota=0\ .
\end{array}
\end{equation}
Altogether implies that only three independent parameters are required to describe the rotation between the physical states ($\eta$, $\eta^\prime$ and $\iota$) and the orthonormal mathematical states
($\eta_q$, $\eta_s$ and $G$).
The rotation matrix can be written in terms of three mixing angles,
$\phi_P$, $\phi_{\eta G}$ and $\phi_{\eta^\prime G}$,
which would correspond to the three Euler angles for a rotation in real, three-dimensional space.
Explicitly,
\begin{equation}
\label{rotation3D}
{\footnotesize
\left(\begin{array}{c}
\eta \\ \eta^\prime \\ \iota
\end{array}\right)=
\left(\begin{array}{ccc}
c\phi_P c\phi_{\eta G} & -s\phi_P c\phi_{\eta G} & -s\phi_{\eta G} \\
s\phi_P c\phi_{\eta^\prime G}-
c\phi_P s\phi_{\eta^\prime G} s\phi_{\eta G} &
c\phi_P c\phi_{\eta^\prime G}+
s\phi_P s\phi_{\eta^\prime G} s\phi_{\eta G} & -s\phi_{\eta^\prime G} c\phi_{\eta G} \\
s\phi_P s\phi_{\eta^\prime G}+
c\phi_P c\phi_{\eta^\prime G} s\phi_{\eta G} &
c\phi_P s\phi_{\eta^\prime G}-
s\phi_P c\phi_{\eta^\prime G} s\phi_{\eta G} &
c\phi_{\eta^\prime G} c\phi_{\eta G}
\end{array}\right)
\left(\begin{array}{c}
\eta_q \\ \eta_s \\ G
\end{array}\right)\ ,
}
\end{equation}
with $(c,s)\equiv (\cos,\sin)$.
In the limit in which $\phi_{\eta G}=\phi_{\eta^\prime G}=0$, the gluonium decouples and the former matrix reduces to the usual rotation matrix describing $\eta$-$\eta^\prime$ mixing,
\begin{equation}
\label{rotation2D}
\left(\begin{array}{c}
\eta \\ \eta^\prime
\end{array}\right)=
\left(\begin{array}{lr}
\cos\phi_P & -\sin\phi_P \\
\sin\phi_P  &  \cos\phi_P
\end{array}\right)
\left(\begin{array}{c}
\eta_q \\ \eta_s
\end{array}\right)\ ,
\end{equation}
where $\phi_P$ is the $\eta$-$\eta^\prime$ mixing angle in the quark-flavour basis related to its octet-singlet basis analog through
$\theta_P=\phi_P-\arctan\sqrt{2}\simeq\phi_P-54.7^\circ$.
An interesting situation occurs when the gluonium content of the $\eta$ meson is assumed to vanish, \emph{i.e.~}$\phi_{\eta G}=0$.
In this particular case,
\begin{equation}
\label{gluoniumetap}
\begin{array}{lll}
X_\eta=\cos\phi_P \ , & Y_\eta=-\sin\phi_P\ , & Z_\eta=0 \ ,\\[1ex]
X_{\eta^\prime}=\sin\phi_P\cos\phi_{\eta^\prime G} \ , &
Y_{\eta^\prime}=\cos\phi_P\cos\phi_{\eta^\prime G} \ , & Z_{\eta^\prime}=-\sin\phi_{\eta^\prime G} \ .
\end{array}
\end{equation}

\end{document}